\documentclass[lettersize,journal]{IEEEtran}
\usepackage{amsmath,amsfonts}
\usepackage{algorithmic}
\usepackage{algorithm}
\usepackage{array}
\usepackage[caption=false,font=normalsize,labelfont=sf,textfont=sf]{subfig}
\usepackage{textcomp}
\usepackage{stfloats}
\usepackage{url}
\usepackage{verbatim}
\usepackage{graphicx}
\usepackage{cite}
\usepackage{amssymb}
\usepackage{color}
\begin{document}

\title{Multi-channel Time Series Decomposition Network \\For Generalizable Sensor-Based Activity Recognition}

\author{Jianguo Pan, Zhengxin Hu, Lingdun Zhang, Xia Cai
\thanks{The authors are with the College of Information, Mechanical and Electrical Engineering, Shanghai Normal University, Shanghai 200234, China (\textit{E-mail of corresponding author: xcai17@fudan.edu.cn}).}}%


\maketitle

\begin{abstract}
Sensor-based human activity recognition is important in daily scenarios such as smart healthcare and homes due to its non-intrusive privacy and low cost advantages, but the problem of out-of-domain generalization caused by differences in focusing individuals and operating environments can lead to significant accuracy degradation on cross-person behavior recognition due to the inconsistent distributions of training and test data. To address the above problems, this paper proposes a new method, Multi-channel Time Series Decomposition Network (MTSDNet). Firstly, MTSDNet decomposes the original signal into a combination of multiple polynomials and trigonometric functions by the trainable parameterized temporal decomposition to learn the low-rank representation of the original signal for improving the extraterritorial generalization ability of the model. Then, the different components obtained by the decomposition are classified layer by layer and the layer attention is used to aggregate components to obtain the final classification result. Extensive evaluation on DSADS, OPPORTUNITY, PAMAP2, UCIHAR and UniMib public datasets shows the advantages in predicting accuracy and stability of our method compared with other competing strategies, including the state-of-the-art ones. And the visualization is conducted to reveal MTSDNet's interpretability and layer-by-layer characteristics.

\end{abstract}



\section{Introduction}
\IEEEPARstart{H}{uman} activity recognition (HAR) is a research hotspot in the field of machine learning (ML) and pattern recognition for the goal of classifying the activities conducted by the participants with ML algorithms~\cite{RN32}, which has been widely used in medical diagnostic monitoring~\cite{RN20,RN21}, human-computer interaction~\cite{RN24,RN25}, identity recognition~\cite{RN22,RN23} and other fields. Compared with video or image-based HAR~\cite{RN26,RN27}, sensor-based HAR has more advantages in cost and privacy. Traditional HAR method based on ML such as SVM~\cite{RN68}, KNN~\cite{RN69} usually concentrates on feature engineering and needs to be redesigned when the recognition task changes. Recently, deep learning methods such as CNN, LSTM~\cite{RN70}, DeepConvLSTM~\cite{RN12} and Transformer~\cite{RN50}, can achieve higher prediction accuracy and have gradually become the main methods for HAR tasks. These models work well when the data of training and test data has the similar or overlapping data distributions, which are achieved by splitting data of each participant into both training and test sets. Nevertheless, this approach proves unwieldy in real-world scenarios, prompting a more pragmatic alternative in the form of cross-person HAR, which entails segregating the data according to different volunteers. Specifically, in the sensor-based cross-person HAR, sensor data exhibits the variations in age, scene and etc., where the distribution of training and test sets fails to conform to the assumption of independent and identical distribution (I.I.D). For example, in the fall detection task, the data of the elderly is difficult to be collected which is often needs to be replaced by the data of the young for model learning. Similarly, in the conventional activity recognition task, the training set comprises data acquired from a controlled laboratory environment, whereas the test set comprises data from complex, real-world environments that exhibit higher levels of environmental noise and activity complexity, exceeding those encountered during training. Remarkably, when confronted with data derived from unobserved subjects, the researchers found that the model would experience a decrease in accuracy and identified some of the reasons as differences in hardware or wearing~\cite{RN28,RN19}.

To deal with the above discrepancies of data distribution, the methods of transfer learning, domain adaptation and domain generalization are presented in the literature~\cite{RN1,RN29}. Transfer learning method usually adopts two steps: training and fine-tuning, where the model is trained with multiple source domains and then is fine-tuned in order to adapt to the target domain. In the fine-tuning process, the auxiliary information of the target domain is more or less introduced~\cite{RN75,RN78} to avoid negative transfer problems. When the data distribution of the target domain is known or unlabeled target domain data is provided, the models of domain adaptation can map the source domain and target domain to the subspace under the same distribution for reducing the domain differences. Most methods of domain adaptation are designed for image data and are incompatible with time series data~\cite{RN76,RN77}. Meanwhile, if there are multiple target domains, transfer learning and domain adaptation methods both need to be re-trained for each target domain. Domain generalization method focuses on obtaining invariant features or domain-independent features to ensure the generalization ability on invisible target domain and reduce the interference of domain-specific features~\cite{RN72}, which does not require any additional information of target domain, compared with transfer learning and domain adaptation.

In this work, we focus on the cross-person sensor-based HAR and treat the data of different volunteers as a single domain. We investigate the error characteristics of sensor data and propose a new model called MTSDNet to hierarchically process the original signal based on the time series decomposition and the design in NBEATS method~\cite{RN2}. Each layer of MTSDNet consists of a decomposer and a classifier, where the decomposer is used to decompose the input signal into specific components and the classifier outputs the classification results of the components. MTSDNet expresses the original signal as a combination of multiple polynomials and trigonometric functions with trainable parameters. 
By using $1 \times 1$ convolution instead of fully connected layers for parameter sharing, MTSDNet applies a single channel temporal decomposition method to multi-channel temporal data.
Meanwhile the normalization within a sliding window is employed at each layer to constrain the feature range for learning the low-rank representation of the original signal. Through these steps, MTSDNet transforms the original distribution into multiple distributions with different constraints so that the distribution of components in the same layer remains constant, while that of different layers varies. This structure allows the model to focus on the component with the smallest distribution difference and improve the model's generalization ability. Our experimental evaluations on DSADS, PAMAP2, OPPORTUNITY, UCIHAR and UniMib datasets validate that our model can outperform state-of-the-art methods with enhanced generalization ability and is suitable for practical HAR applications. 

In sum, the main innovations of MTSDNet are as follows:

\begin{enumerate}
\item{A novel decomposer structure including the normalization and denormalization within a sliding window to constrain the layer distribution is designed for classification, and the statistical information is integrated into the features through multi-view approach to recover the information lost during normalization.}
\item{A novel converter is presented that can directly convert the additive model of time series decomposition into multiplicative model to adapt to different signals.}
\item{A multi-channel temporal decomposition structure is proposed, which replaces the fully connected layer with $1 \times 1$ convolution to expand single-channel temporal decomposition into multi-channel temporal decomposition.}
\end{enumerate}

\begin{figure*}[!t]
\centering
\includegraphics{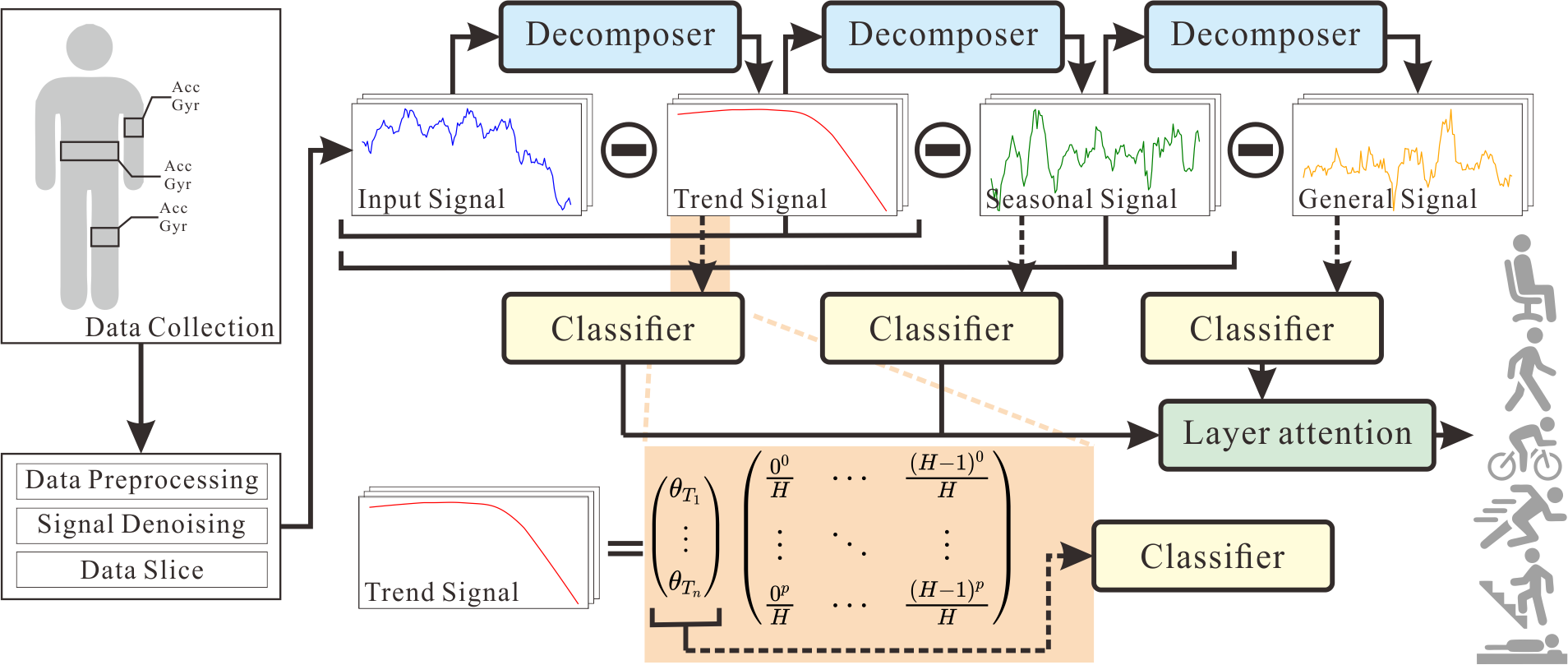}
\caption{Illustration of the proposed method. Firstly, data is collected through sensors such as accelerometers and gyroscopes. After data preprocessing, denoising and slicing, input data is decomposed into multiple components by decomposer. Each component can be considered as a multiplication combination of preterm coefficients and constraint terms. The orange part indicates that the classifier use the preterm coefficients of each component. Finally, MTSDNet integrates classification results using a layer attention. Here it adopts the structure of the model MTSDNet-A-tsg, which is an additive model and decompose the signal to one trend, one seasonal and one general component. More details can be found in Section~\ref{Overall Structure}.}
\label{Fig_all}
\end{figure*}

The proposed MTSDNet is shown in Figure~\ref{Fig_all}. As the name suggests, the input signal is decomposed into specific forms by MTSDNet. MTSDNet predicts each component separately and the attention is used to obtain the final result.

\section{Related Work}
Feature learning method for sensor-based HAR can roughly be grouped into two categories: classical machine learning and deep learning. The general processing steps of machine learning mainly consists of using statistical features or fourier transform for feature engineering and using KNN~\cite{RN79}, XGBoost~\cite{RN80} and other algorithms for classification. Time series decomposition method, which takes a compositional perspective on time series data, is often used in conjunction with traditional ML methods for sensor-based HAR. Recently, this method has also been incorporated into deep learning approaches. In this regard, a novel recognition method based on online feature vector calculation and multi-way decomposition algorithm has been proposed~\cite{RN73}. FuzzyAct jointly classifies and detects human activities through Discrete Wavelet Transform (DWT) and Recurrent Neural Network (RNN)~\cite{RN74}. Distribution-Embedded Deep Neural Network (DDNN) integrates data features from multiple modalities and enhances the model's generalization ability by fusing spatial, temporal and statistical domain information~\cite{RN30}. DeepConvLSTM~\cite{RN12} effectively applies convolutional network and Long Short-Term Memory (LSTM) network  to different sensor modes. Attention mechanism is introduced later for further enhancing the model's feature extraction ability~\cite{RN46}. These methods are characterized by enhancing the model's feature extraction ability and improving generalization ability to some extent.

Looking at cross-person HAR from the perspective of domain generalization, many methods refer to the training and fine-tuning steps in transfer learning, which train on the source domain and fine-tune on the partial data from target domain. However, in sensor-based HAR tasks, due to the differences in the number and location of sensors, as well as the lack of sensor data, it is not feasible to simply employ transfer learning method. Domain-Adversarial Training of Neural Networks (DANN) achieves the error minimization of task classification and the error maximization of domain classification by adopting the methods of gradient reversal and multi-task learning, thereby forcing the model to focus only on the domain-irrelevant part during feature extraction~\cite{DANN}. Convolutional deep Domain Adaptation model for Time Series data (CoDATS) constructs a domain adaptive weakly supervised method and uses target domain data for domain adaptation to improve the model's accuracy~\cite{RN29}. Generalizable Independent Latent Excitation (GILE) adopts multi-task learning and source domain-independent architecture to separate volunteer-specific features and volunteer-independent features based on volunteer source labels so that only volunteer-independent features are used for predictions to enhance cross-person generalization ability~\cite{RN1}.

\section{Method}
\subsection{Error Analysis}\label{Error Analysis}
Regarding the analysis of the sensor data used for cross-person HAR, it is found that the following factors are the main reasons for the distribution differences between different volunteers:

\begin{enumerate}
\item{Data bias: arising from the wearing differences of the sensors in the same position and the variations in the three-axis direction caused by the sensor production, this bias is manifested in the original signal as static or background deviation. Taking the accelerometer as an example, the bias appears as variations in data direction and amplitude, contingent upon a variety of factors, including sensor placement, rotation, motion-induced displacement and so on.}
\item{Volunteer activity differences: mainly derived from the signal differences reflected on sensors under the amplitude, frequency and composite of dynamic actions between different volunteers, as well as signal differences of complex actions on the sensors. This difference is manifested in the original signal as features such as the amplitude, frequency and periodicity of seasonal signals.}
\item{Secondary noise: mainly arising from the relative movement of sensors to the body caused by human motion when the sensors and the body are not tightly attached. The information collected by the sensors contains the movement of the sensors relative to the body in addition to the human motion, which is manifested as data bias changing over time in other original signal.}
\end{enumerate}
For the above, data bias is rarely singled out and addressed specifically. The work~\cite{RN19} explains the data bias problem caused by the sensor hardware, but does not consider the data bias problem caused by the fixed way of the sensor on human body. 
To address this, we propose a time series decomposition approach to separate signals which causes distribution differences. The original time series data is decomposed into the trend, seasonal and residual items. The data bias can be captured by the trend item, the periodicity of human activity can be captured by the seasonal item, and the remaining part forms the residual item. The construction of this method originates from traditional temporal decomposition methods and explicit feature alignment in domain generalization. The former plays an important role in many time series decomposition tasks, which can separate signal components with partial interpretability. The latter mainly focuses on mapping features to feature spaces with similar or identical features to alleviate or eliminate feature distribution differences. By learning a decomposition, each layer of components is located in the same feature space. Considering the fact that human behavior may not have the periodicity with constant frequency, a general term is added to the decomposition structure to extract the non-periodic features.

\subsection{Problem Statement}
For cross-person HAR, we consider each volunteer as a domain which can be defined as a joint distribution $\mathbb{P}^{d}(x,y)$ on $\mathcal{X}\times\mathcal{Y}$, where $\mathcal{X}$, $\mathcal{Y}$ and $d \in \mathcal{D} = \{1,...,D\}$ denote activity instance space, activity class space and the index of source domain. The difference between the cross-person HAR and traditional HAR during modeling is the division way of training and test set. Training set can be defined as $\{(X^{d},y^{d})\thicksim \mathbb{P}^{d}(x,y)_{d=i}^{D}\}$ and test set can be defined as $\{(X^{d},y^{d})\thicksim \mathbb{P}^{d}(x,y)_{d=j}^{D}\},\forall i\neq j $ where $i,j\in \mathcal{D}$. In this task, training set and test set have the same activity class space. The goal of cross-person HAR is to train a generalizable model within training set, which is able to generalize well on test set and is unaffected by volunteer differences.

To simplify the task, a sliding window is used to divide the data into fixed-length sequence which can be defined as $\{(F_{k}^{t},y)$. $F_{k}^{t}\in\mathbb{R}^{k\times t}\}$. Here $k$ is the number of sensor channels, $t$ is the sliding window length, $y$ is the activity label of the last moment in the sequence. Combined with Section~\ref{Error Analysis}, the model decomposes the original data into multiple components, which can be expressed as either additive model or multiplicative model. For example, in the additive model shown as
$F_{k}^t=\sum_{i=1}^q p_i+I$, 
where $p_i\in\mathbb{R}^{k\times t}$ is the component, $q$ means the number of components and $I$ is the noise. After decomposition, each component is classified by the designed model and the final classification result can be obtained by a weighted sum of the outputs from different components. The result is shown as
$\hat{y}=\sum_{i=1}^q (w_i\times f_i(p_i))$, 
where $w_i$ is the weight and $f_i$ is the classification model for each component.

\subsection{Overall Structure}
\label{Overall Structure}
\begin{figure}[!t]
\centering
\includegraphics[width=3in]{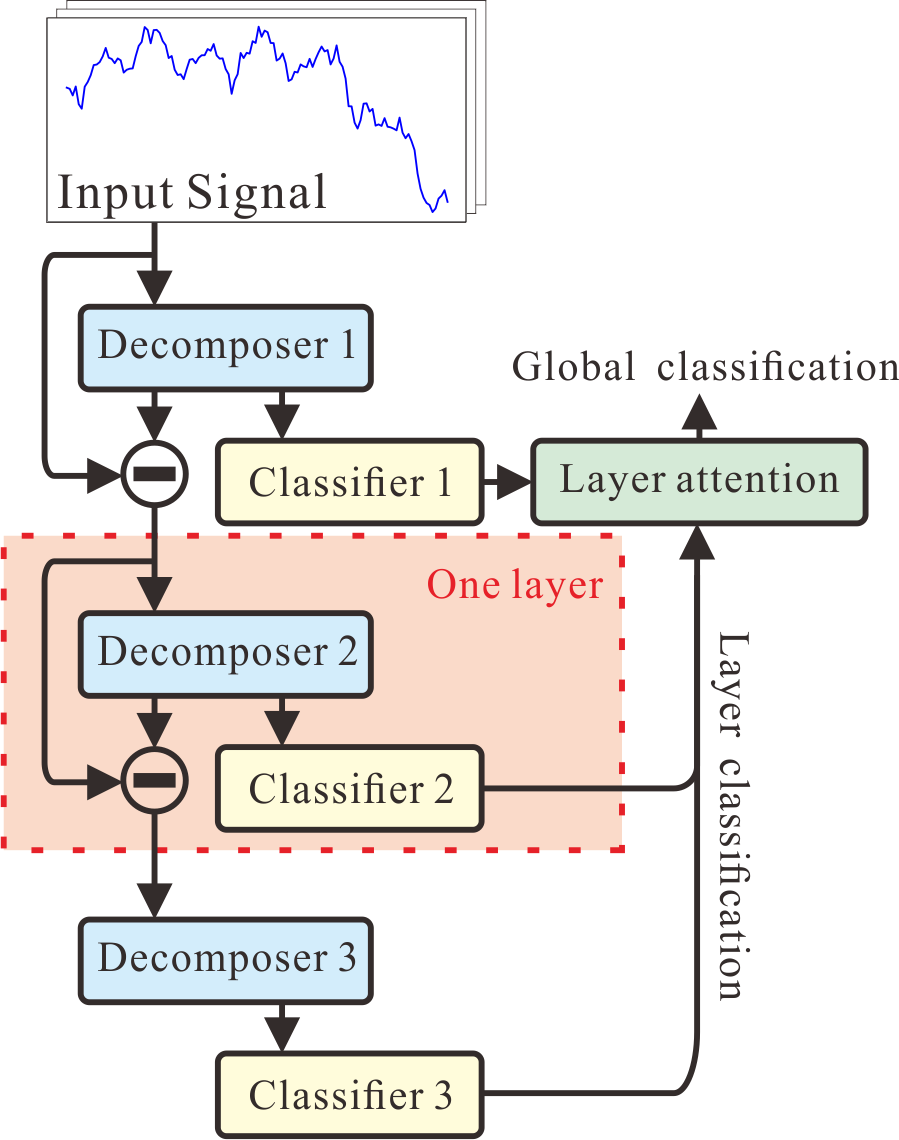}
\caption{The overall structure of proposed MTSDNet according to the additive model with three layers and no decomposition term output in the last layer. The model input is the sensor data divided into many sliding windows and the output is a weighted classification results of multiple layers.}
\label{MTSDNet}
\end{figure}

The model framework of MTSDNet is shown in Figure~\ref{MTSDNet} , which is composed of multiple layers with each layer having one input and two outputs. The left output is the decomposed signal and the right is the classification output of the layer, where the classification outputs of different layers are aggregated into a final result through the layer attention mechanism. The input signal is processed by subtraction or division in the multi-layer block from top to bottom to form an additive or multiplicative model. In Figure~\ref{MTSDNet}, the symbol '-' shows additive model. Each layer consists of a decomposer and a classifier, which are responsible for time series decomposition and classification of component features, respectively. In light of the fact that traditional time series decomposition methods are characterized by both additive and multiplicative model, MTSDNet preserves consistency with these methods by adopting the same designs. And this framework enables the decomposition of the original data into additive or multiplicative forms subject to the constraints of MTSDNet shown as following:
\begin{equation}
\label{deqn_1}
S_{ori}=\sum_{i=1}^n T_i+\sum_{i=1}^m S_i+\sum_{i=1}^o G_i+I
\end{equation}
\begin{equation}
\label{deqn_2}
S_{ori}=\prod_{i=1}^n T_i\times\prod_{i=1}^m S_i\times\prod_{i=1}^o G_i\times I
\end{equation}
where $S_{ori}$ is the original time series signal. $T$, $S$, $G$ and $I$ are the trend, seasonal, general and noise items. $n$, $m$, $o$ are the number of $T$, $S$, $G$. The polynomial constraint and trigonometric function constraint are used for trend and seasonal items, respectively. The general item consists of two $1\times 1$ convolutional layers without any constraint. The remaining part after the completion of temporal decomposition is $I$, which is not used for classification. The items of $T$, $S$ and $G$ can be decomposed step by step with multiple layers. The $T$ trend item uses the following polynomial constraint:
\begin{equation}
\label{deqn_3}
{T}_i=\theta_T T^{\prime}
\end{equation}
\begin{equation}
\label{deqn_T}
T^{\prime}=\begin{pmatrix}\frac{0^0}{H} & \cdots & \frac{(H-1)^0}{H}  
\\ \vdots & \ddots & \vdots  
\\ \frac{0^p}{H} & \ldots & \frac{({H-1})^p}{H} \end{pmatrix}
\end{equation}
where $T'$ is a matrix representing polynomial constraints, $\theta_T$ is the feature generated through trend decomposer, $p$ is the number of polynomials, and $H$ is the length of the sliding window. Polynomial constraints are expressed in matrix form and the decomposition form of the signal is obtained by matrix multiplication with the computed high-dimensional features. This is equivalent to restricting the trend item to a specific structured polynomial function and allowing the model to learn the coefficients of each polynomial term. The $S$ trend item uses the following trigonometric constraint:

\begin{equation}
\label{deqn_4}
{S}_i=\theta_S S^{\prime}
\end{equation}
\begin{equation}
\label{deqn_S}
S^{\prime}=\begin{pmatrix}
\cos(\frac{2\pi\times0\times0}{H}) & \cdots & \cos[\frac{2\pi\times0\times(H-1)}{H}] 
\\ \vdots & \ddots & \vdots 
\\ \cos(\frac{2\pi\times(\frac{H}{2}-1)\times0}{H}) & \cdots & \cos[\frac{2\pi\times(\frac{H}{2}-1)\times(H-1)}{H}] 

\\ \sin(\frac{2\pi\times0\times0}{H}) & \cdots & \sin[\frac{2\pi\times0\times(H-1)}{H}] 
\\ \vdots & \ddots & \vdots 
\\ \sin(\frac{2\pi\times(\frac{H}{2}-1)\times0}{H}) & \cdots & \sin[\frac{2\pi\times(\frac{H}{2}-1)\times(H-1)}{H}] 

\end{pmatrix}
\end{equation}
where $S'$ is a trigonometric constraint expressed in matrix form, and $\theta_S$ is the feature generated through seasonal decomposer. Similar to the polynomial function constraint, trigonometric function constraint can also be expressed as a restriction on the seasonal term, which is a combination of specific trigonometric functions with coefficients that are learned during the training process.

By the above construction using $T'$ and $S'$, the decomposer can decompose input signal into the product of coefficient matrix and temporal characteristic matrix to express the temporal characteristics of the input signal. According to the constraint property, the input signal will have a high coefficient on part of features, while the rest of coefficients are close to zero, thus constituting a low-rank feature. 

\subsection{Decomposer Structure}
\begin{figure}[!t]
\centering
\includegraphics[width=2.8in]{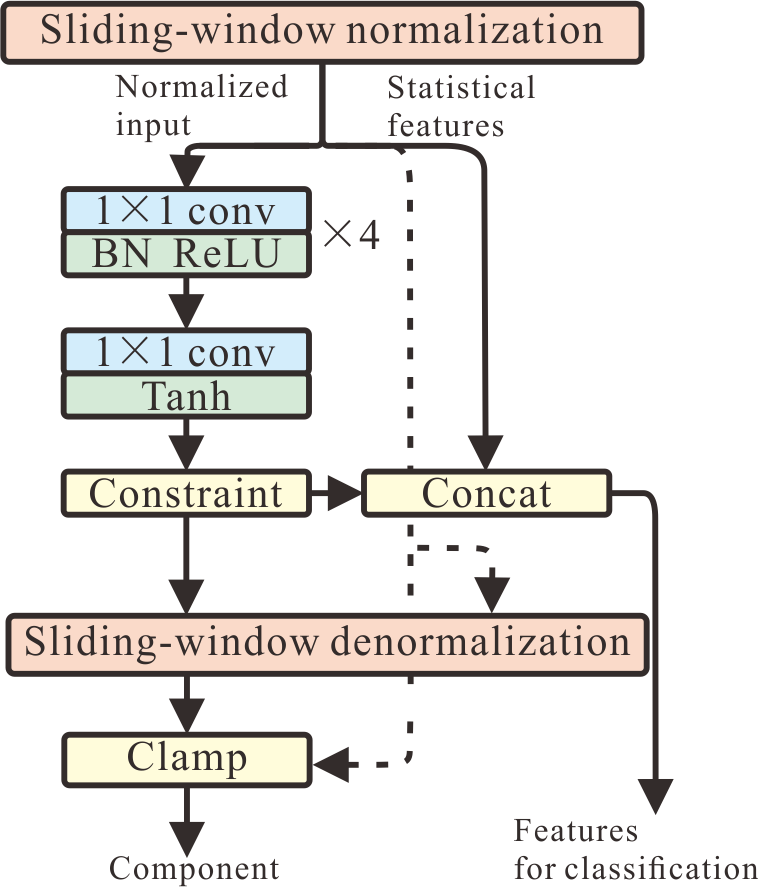}
\caption{The illustration of decomposer structure with one input and two outputs. The left and right sides output the decomposition term and  classification results, respectively. The dashed line indicates that the data is used as parameter for the function instead of function input.}
\label{Decomposer}
\end{figure}

The decomposer is composed of a sliding window normalization and denormalization module, a statistical eigenmodel fusion module, a constraint control module and a distribution clamping module. The structure of the decomposer is illustrated in Figure~\ref{Decomposer}. The flow of the decomposer is described as following steps, with the solid line representing data utilized as features and the dashed line indicating data utilized for parameter input.

\begin{enumerate}
\item{The input signal of current layer is normalized within a sliding window where the statistical information (minimum, mean, maximum and variance) can be obtained. In addition, a fixed bias of 1e$^{-5}$ is added to the variance in case of zero variance.}
\item{Four $1 \times 1$ convolutional layers, batch normalization (BN) and ReLU are employed for channel-by-channel information extraction with weight sharing. Then a $1\times1$ convolutional layer is used to obtain the fixed-size features with Tanh activation function, which is multiplied with the constraint term in order to extract the signal.}
\item{The extracted components in the time domain are obtained through multiplying the high-dimensional features by the constraint terms, and then the statistical information within the sliding window is used for denormalization. Subsequently, the signal is corrected according to the additive or multiplicative model of the temporal decomposition. Finally, the minimum and maximum values from the statistical information are employed for clamping. The range of the signal for the additive model is limited to 1.1 times the range of the input data, while the multiplicative model is constrained to a range of (0.5, 2). The range clamping for the additive model is specifically calculated as follows:
\begin{equation}
\label{deqn_5}
\begin{split}
D_{out}=clamp(\min(D_{in})-C\times|\min(D_{in})|,
\\ \max(D_{in})+C\times|\max(D_{in})|,D_{in})
\end{split}
\end{equation}
where $D_{in}$ and $D_{out}$ are the input and output feature of the clamp module, respectively. $C$ is constraint weights to control the range of component. In additive model, the value of C is set to 0.1. The clamp module can restrict the data within a specific range.
}
\item{The classifier of the model is constructed by three-layer Multilayer Perceptron (MLP). Considering that the normalization in the first step can lose statistical information of input features, such as mean and variance, which contains behaviorally relevant features and should not be ignored, both high-dimensional features and statistical features are concatenated together as the inputs of classifier for the classification.}
\end{enumerate}

Through the above steps, MTSDNet is finally designed based on the framework of additive model. Then we further propose that MTSDNet can be transformed into a multiplicative one to fully utilize input signals using the following expression:
\begin{equation}
\label{deqn_6}
D_*=Relu(D_+)+(\dfrac{1}{Relu((-1)\times D_+)+1})
\end{equation}
where $D_+$ is the feature extracted from MTSDNet under the additive model, and $D_*$ is the transformed features for multiplicative model. This operation is located between sliding-window denormalization and clamp.

\section{Experiments}
\subsection{Experimental Environment}
All of the experiments are conducted on a sever containing NVIDIA GeForce RTX 3080 with pytorch-lightning and pytorch, which also are executed with the same training hyperparameters, including a batch-size of 512, a learning rate of 3e$^{-3}$, 20 training epochs and 10 repetitions of each model with random seeds. We follow the assessment approach of domain generalization during the experiments. Specifically, the data from different volunteers is treated as the different domains and the data is divided into the training and test set using single-domain or multi-domain splitting method. The mean and variance of accuracy under 10 training sessions are adopted as evaluation metrics. 

For the comparison of experimental results, a variety of traditional methods including the commonly used LSTM, GRU and three-layer MLP, as well as DeepConvLSTM and Transformer are provided. DANN and GILE are selected for comparing model performance under domain generalization. Although DANN is proposed as a domain adaptation method, it is changed as a domain generalization method to uses the labels between different source domains as auxiliary targets for unifying measurement in this work. GILE uses 2 and 0.5 as loss hyperparameters in multi-task learning.

\subsection{Experimental Dataset}

Five HAR benchmark datasets are used for the experimental evaluation in this paper, including the DSADS~\cite{RN14}, PAMAP2~\cite{RN15}, OPPORTUNITY~\cite{RN65}, UCIHAR~\cite{RN81} and UniMib~\cite{RN82} datasets. The parameters associated with each dataset involved in the experiments are listed in Table~\ref{tab:table1}.

\begin{table*}[!t]
\caption{The parameter summary of involved public datasets\label{tab:table1}}
\centering
\scalebox{0.94}{
\begin{tabular}{| c | c | c | c | c | c |}
\hline
Parameters & UCIHAR & PAMAP2 & DSADS & OPPORTUNITY & UniMib \\
\hline
Volunteer Num & 30 & 9 & 8 & 4 & 30 \\
\hline
Sensor Channel Num & 9 & 54 & 45 & 113 & 3 \\
\hline
Activity Num & 6 & 18 & 19 & 18 & 2 \\
\hline
Preprocessing & Standard & Standard & Standard & MinMax & Standard \\
\hline
Length & 128 & 128 & 32 & 64 & 151 \\
\hline
Stride & 32 & 32 & 28 & 32 & 151 \\
\hline
Split & 1-7,8-14,15-21,22-28 & 1,2,3,4,5,6,7,8 & 1,2,3,4,5,6,7,8 & 1,2,3,4 & 1-10,11-20,21-30 \\
\hline
Label-unbalance & Null & Null & Null & Weighted CE LOSS & Null \\
\hline
\end{tabular}}
\end{table*}

The `Part' in the paper contains either one or several volunteers generated from dataset by different volunteer splitting methods. For example, `Part2' means the second part of volunteer data is taken as test set while the rest part is used as training set and then experiment is conducted for 10 times with random seeds. 
In addition, the experiments focus on gesture recognition task of OPPORTUNITY dataset and fall detection task of UniMib dataset. The 9th volunteer on PAMAP2 dataset and the 29th and 30th volunteers on UCIHAR dataset are only used for training due to limited data on corresponding volunteers. MTSDNet can be designed several structural versions, such as MTSDNet-A-stg, MTSDNet-M-g3 or MTSDNet-A-t3s3g3, where A and M mean MTSDNet is based on additive and multiplicative model, respectively. t3s3g3 means that the decomposition structure of the model is 3 trend blocks, 3 seasonal blocks and 3 general blocks. If no number is attached to the letter, it means one block. In this paper, we refer to the feature order obtained by traditional temporal decomposition and choose 4 versions of MTSDNet: MTSDNet-A-tsg, MTSDNet-M-tsg, MTSDNet-A-t3s3g3 and MTSDNet-M-t3s3g3, to highlight the differences between the additive and multiplicative models and show the effect of different stacked ways.

\subsection{Experimental results}

\begin{table*}[!t]
\caption{Comparison of prediction accuracy achieved by various methods on five datasets. The best result on each dataset is highlighted in bold.\label{tab:table2}}
\centering
\begin{tabular}{| c | c | c | c | c | c |}
\hline
Methods & UCIHAR & PAMAP2 & DSADS & OPPORTUNITY & UniMib \\
\hline
GRU & 91.79 & 64.96 & 81.87 & 44.07 & 94.85 \\ \hline
LSTM & 90.97 & 64.65 & 81.23 & 21.87 & 89.48  \\ \hline
MLP & 90.52 & 64.61 & 79.47 & 68.50 & 96.29  \\ \hline
Transformer & 90.57 & 66.20 & 84.29 & 47.47 & 95.77  \\ \hline
DeepConvLSTM & 92.16 & 69.86 & 85.02 & 54.73 & 94.94  \\ \hline
MTSDNet-A-tsg & 91.67 & 71.92 & 91.28 & \textbf{78.63} & \textbf{98.92}  \\ \hline
MTSDNet-A-t3s3g3 & 91.70 & \textbf{73.34} & \textbf{92.16} & 78.14 & 98.85  \\ \hline
MTSDNet-M-tsg & 92.32 & 72.91 & 91.28 & 78.30 & 98.81  \\ \hline
MTSDNet-M-t3s3g3 & \textbf{92.42} & 72.50 & 90.90 & 74.58 & 98.74  \\ \hline
DANN & 89.29 & 63.30 & 81.30 & 55.49 & 94.75  \\ \hline
GILE & 92.26 & 72.37 & 87.49 & 58.32 & 97.00  \\ \hline
\end{tabular}
\end{table*}

\begin{table*}[!t]
\caption{Comparison of parameters achieved by various methods on five datasets.\label{tab:table_a}}
\centering
\begin{tabular}{| c | c | c | c | c | c |}
\hline
Methods & UCIHAR & PAMAP2 & DSADS & OPPORTUNITY & UniMib \\
\hline
GRU & 54.1 & 65.4 & 69.7 & 95.6 & 51.3 \\ \hline
LSTM & 71.9 & 86.7 & 92.1 & 126 & 68.4  \\ \hline
MLP & 165 & 608 & 203 & 945 & 75.4  \\ \hline
Transformer & 437 & 467 & 434 & 453 & 422  \\ \hline
DeepConvLSTM & 86.8 & 222 & 222 & 518 & 49.4  \\ \hline
MTSDNet-A-tsg & 51.4 & 83.2 & 75.4 & 159 & 64.7  \\ \hline
MTSDNet-A-t3s3g3 & 154 & 249 & 226 & 479 & 145  \\ \hline
DANN & 216 & 659 & 255 & 996 & 126  \\ \hline
GILE & 225 & 282 & 264 & 415 & 221  \\ \hline
\end{tabular}
\end{table*}

The contrast accuracy results achieved by different methods are listed in Table~\ref{tab:table2}, which are the average prediction accuracy through traversing all Parts as the test set on the five datasets
in Table~\ref{tab:table1}. 
As shown in Table~\ref{tab:table2}, MTSDNet and GILE achieve better accuracy compared with other methods, where the average accuracy of MTSDNet surpasses all existing practice on all datasets and GILE is ranked the 2nd on UCIHAR, PAMAP2, DSADS and UniMib datasets. And on UCIHAR and PAMAP2 datasets, the gap of average accuracy between MTSDNet and GILE is small within 1\% or less. Specifically, the average improvement of MTSDNet is about 9\% on the DSADS dataset, 28\% on the OPPORTUNITY dataset, 4\% on the UniMib dataset, 6\% on PAMAP2, and 1\% on the UCIHAR dataset. 

When the dataset is imbalanced, accuracy may overestimate the classification ability of the model and precision, recall, or F1-score for evaluation is more accurate. We also evaluated these metrics and the detailed results is shown in Table S1-15. The performance of precision, recall and F1-score is consistent with the accuracy. Under F1-score evaluation, MTSDNet has advantage on DSADS, PAMAP2, OPPORTUNITY and UniMib datasets. Specifically, the average improvement of MTSDNet on F1-score is about 10\% on the DSADS dataset, 20\% on the OPPORTUNITY dataset, 5\% on the UniMib dataset, 5\% on the PAMAP2 dataset and 1\% on the UCIHAR dataset.

In addition, the DANN method is less effective than MLP when the target domain is not visible for the model, which shows that DANN is not suitable for domain generalization because negative transfer phenomenon occurred in this method. As for the multi-layer stacking of MTSDNet, it is shown that for the additive model, the multi-layer stacking can achieve higher accuracy in some cases, and even if the accuracy is not improved, the decrease of its accuracy is smaller. For the multiplicative model, the accuracy decreases and only on the UCIHAR dataset, MTSDNet-M-t3s3g3 is superior. In the comparison of traditional methods, the accuracy difference of LSTM, GRU, MLP, DeepConvLSTM and Transformer is small on the DSADS, PAMAP2, UCIHAR and UniMib datasets, and DeepConvLSTM performs relatively better. However, on the OPPORTUNITY dataset with severe label imbalance, MLP is the optimal choice in traditional methods.

Table~\ref{tab:table_a} shows the parameters of multiple models on five datasets and the unit of parameter is K. The result ensures that the parameters of MTSDNet-tsg is approximately equal to the smallest model's parameters in the comparing methods and it can exclude the improvement of accuracy caused by the increase in model size. Based on the average accuracy in Table II, it can be seen that the proposed method can achieve better classification accuracy with relatively reasonable model parameters.

\begin{figure*}[!t]
\centering
\includegraphics[width=5.5in]{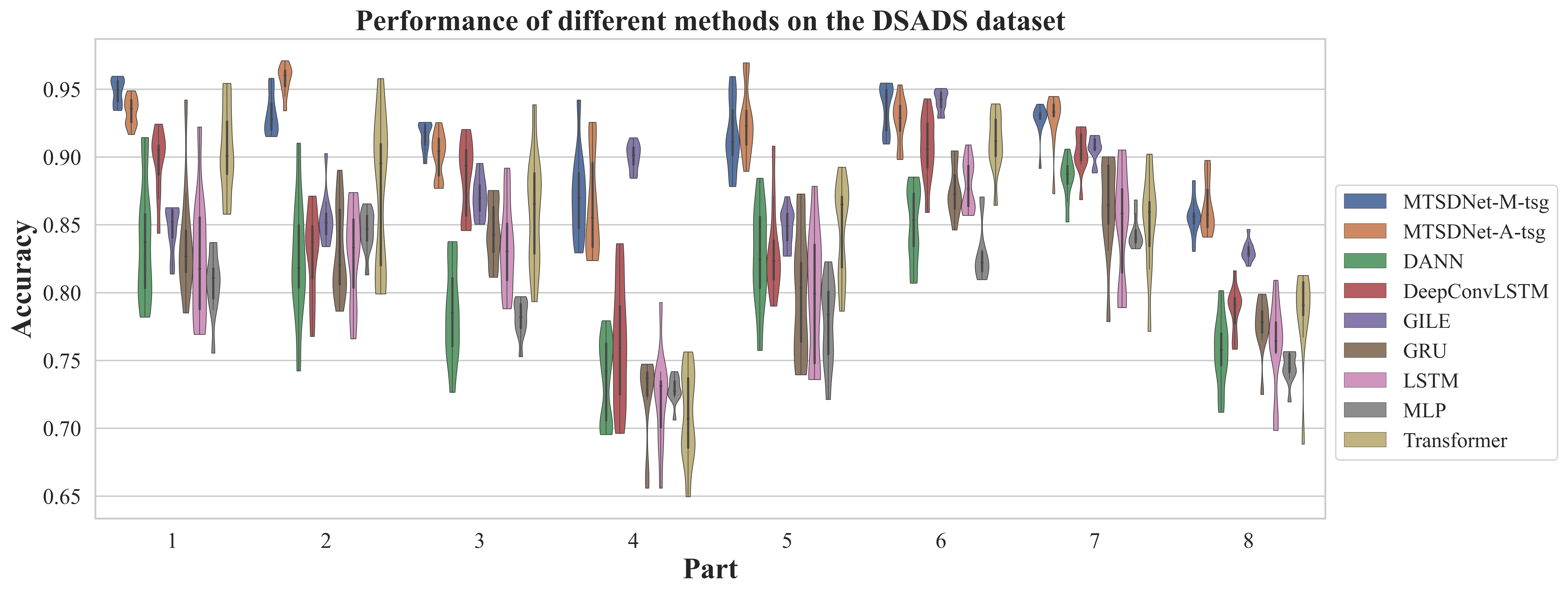}
\caption{Violin plot of accuracy distribution on the DSADS dataset. MTSDNet shown in blue and orange achieves higher accuracy than other baselines on almost all Parts.}
\label{DSADS_acc}
\end{figure*}

\begin{figure*}[!t]
\centering
\includegraphics[width=5.5in]{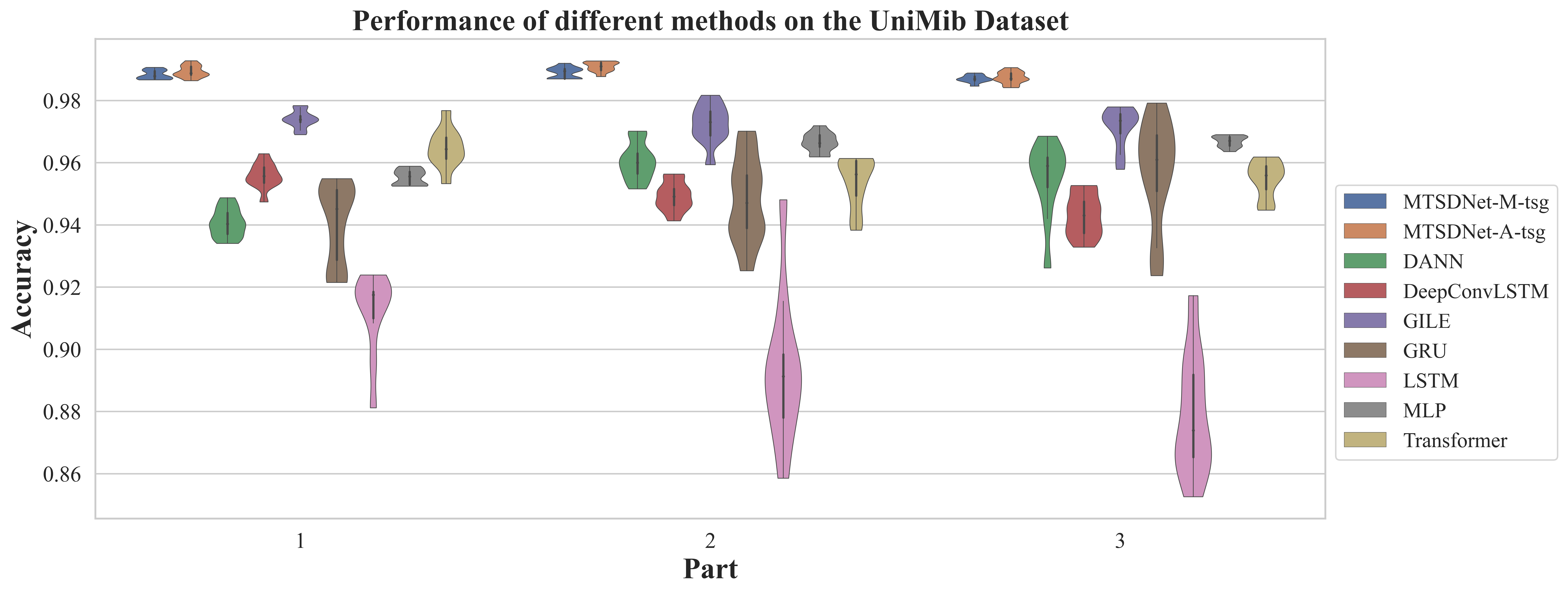}
\caption{Violin plot of accuracy distribution of UniMib dataset. MTSDNet shown in blue and orange achieves higher accuracy than other baselines on 3 parts and also has higher stability.}
\label{UniMib_acc}
\end{figure*}

For the detailed experimental results, violin plots are provided for the DSADS and UniMib datasets to highlight the best and worst cases on each domain as shown in Figure~\ref{DSADS_acc} and Figure~\ref{UniMib_acc}, which present the distribution of classification accuracy obtained after 10 different random seed for training and are represented with equal width in order to highlight view. Compared with the histogram error lines, the violin plots can not only show the average and variance accuracy of the model, but also show the specific distribution characteristics. As shown in Table~\ref{tab:table3} and Figure~\ref{DSADS_acc}, MTSDNet exhibits relatively good stability in all volunteer partitioning on the DSADS dataset (with shortest height of color band) and has the highest average accuracy (with highest color band position), only slightly lower than GILE on Part6. GILE has the second highest average accuracy on the DSADS dataset. On Part4, Part5, Part6 and Part8, GILE has slightly better stability than MTSDNet as shown in Figure~\ref{DSADS_acc}. The average accuracy of traditional methods on the DSADS dataset is about 80\% to 85\%. Although Transformer exhibits optimal accuracy compared to MTSDNet on Part1, Part2 and Part3, its average accuracy is low due to existing worst-case accuracy. On the DSADS dataset, different volunteers have the preferences for the constructing way of MTSDNet by addition or multiplication, where Part1, Part3 and Part4 are more suitable for using multiplicative models. In addition, when using a higher stacked decomposition model, the accuracy of multi-layer stacked MTSDNet-A-t3s3g3 is generally improved by 1\% or more compared to the MTSDNet-A-tsg model, while this phenomenon is not reflected in MTSDNet-M-t3s3g3.

\begin{table*}[!t]
\caption{The accuracy comparison of different methods on the DSADS dataset from Part1 to Part8.\label{tab:table3}}
\centering
\begin{tabular}{|l|l|l|l|l|l|l|l|l|}
\hline
        Method & Part1 & Part2 & Part3 & Part4 & Part5 & Part6 & Part7 & Part8  \\ \hline
        GRU & 83.81(4.38) & 83.11(3.46) & 84.48(2.29) & 72.23(3.27) & 80.02(4.71) & 87.53(1.94) & 86.24(3.74) & 77.54(2.08)  \\ \hline
        LSTM & 82.59(4.70) & 82.89(3.63) & 83.35(3.33) & 72.26(3.67) & 79.70(5.00) & 87.94(1.87) & 85.01(4.21) & 76.08(3.38)  \\ \hline
        MLP & 80.64(2.42) & 84.55(1.55) & 78.10(1.37) & 72.83(1.01) & 77.77(3.21) & 82.95(2.18) & 84.40(1.14) & 74.48(1.14)  \\ \hline
        Transformer & 90.57(3.30) & 87.46(5.75) & 86.07(4.31) & 70.91(3.43) & 84.72(3.64) & 91.04(2.21) & 85.17(3.80) & 78.41(3.66)  \\ \hline
        DeepConvLSTM & 89.71(2.50) & 82.95(3.32) & 88.45(2.78) & 76.14(4.58) & 83.13(3.49) & 90.69(2.49) & 90.37(1.61) & 78.72(1.75)  \\ \hline
        MTSDNet-A-tsg & 93.37(1.05) & 95.71(1.12) & 90.14(1.67) & 86.55(3.87) & \textbf{92.62(2.58)} & 92.73(1.73) & 92.79(2.11) & 86.34(1.99)  \\ \hline
        MTSDNet-A-t3s3g3 & 93.73(1.83) & \textbf{95.84(0.52)} & 90.77(1.16) & 88.90(3.30) & 92.30(1.77) & 94.02(1.14) & \textbf{93.19(0.72)} & \textbf{88.56(1.41)}  \\ \hline
        MTSDNet-M-tsg & 94.81(0.97) & 93.13(1.49) & 91.52(1.04) & 87.23(3.32) & 91.58(2.64) & 93.68(1.74) & 92.69(1.32) & 85.59(1.42)  \\ \hline
        MTSDNet-M-t3s3g3 & \textbf{95.96(0.73)} & 91.81(2.32) & \textbf{94.12(0.93)} & \textbf{91.03(1.87)} & 87.67(2.05) & 92.78(1.29) & 90.55(1.13) & 83.26(1.76)  \\ \hline
        DANN & 84.04(4.63) & 82.45(4.73) & 78.53(3.65) & 73.59(3.18) & 82.57(3.83) & 85.06(2.83) & 88.47(1.57) & 75.71(2.67)  \\ \hline
        GILE & 84.71(1.75) & 85.50(1.96) & 87.08(1.47) & 90.03(0.98) & 84.80(1.44) & \textbf{94.12(0.80)} & 90.59(0.95) & 83.07(0.76) \\ \hline
\end{tabular}
\end{table*}

\begin{table}[ht]
\caption{The accuracy comparison of different methods on the UniMib dataset from Part1 to Part3.\label{tab:table4}}
    \centering
    \begin{tabular}{|l|l|l|l|}
    \hline
        METHOD & Part1 & Part2 & Part3  \\ \hline
        GRU & 94.04(1.32) & 94.79(1.42) & 95.71(1.76)  \\ \hline
        LSTM & 91.19(1.35) & 89.36(2.46) & 87.88(2.00)  \\ \hline
        MLP & 95.53(0.25) & 96.66(0.32) & 96.68(0.18)  \\ \hline
        Transformer & 96.47(0.65) & 95.37(0.83) & 95.48(0.59)  \\ \hline
        DeepConvLSTM & 95.58(0.43) & 94.92(0.48) & 94.31(0.69)  \\ \hline
        MTSDNet-A-tsg & 98.93(0.20) & 99.08(0.16) & \textbf{98.74(0.20)}  \\ \hline
        MTSDNet-A-t3s3g3 & \textbf{99.01(0.16)} & \textbf{99.14(0.25)} & 98.71(0.18)  \\ \hline
        MTSDNet-M-tsg & 98.83(0.15) & 98.90(0.18) & 98.70(0.12)  \\ \hline
        MTSDNet-M-t3s3g3 & 98.60(0.16) & 99.06(0.18) & 98.55(0.17)  \\ \hline
        DANN & 94.07(0.47) & 96.04(0.61) & 95.47(1.25)  \\ \hline
        GILE & 97.39(0.29) & 97.24(0.65) & 97.15(0.66)  \\ \hline
    \end{tabular}
\end{table}

The UniMib dataset contains four tasks, and our experiment mainly focuses on the binary classification of fall detection. As seen from Table~\ref{tab:table4} and Figure~\ref{UniMib_acc}, all methods can achieve an accuracy of over 90\% in this task. Importantly, MTSDNet not only has the best average accuracy and minimum variance, but also achieves an accuracy of nearly 99\%, which is much higher than other methods that generally have the accuracy of around 95\%. As shown in Figure~\ref{UniMib_acc}, the stability of our method is also significantly better than others and GILE method achieves sub-optimal accuracy of 97\%.

\begin{table}[ht]
\caption{The accuracy comparison of different methods on the UCIHAR dataset from Part1 to Part4.\label{tab:table5}}
    \centering
    \resizebox{1\columnwidth}{!}{
    \begin{tabular}{|l|l|l|l|l|}
    \hline
        Method & Part1 & Part2 & Part3 & Part4  \\ \hline
        GRU & \textbf{93.76(0.85)} & 86.08(1.36) & 91.67(0.89) & \textbf{95.65(0.58)}  \\ \hline
        LSTM & 92.21(1.75) & 84.82(1.80) & 91.55(2.30) & 95.30(1.80)  \\ \hline
        MLP & 92.11(0.45) & 85.77(0.76) & 91.83(2.09) & 92.37(0.91)  \\ \hline
        Transformer & 93.00(0.77) & 84.44(1.40) & 91.53(1.96) & 93.34(1.10)  \\ \hline
        DeepConvLSTM & 92.70(0.78) & 88.07(1.02) & 93.09(0.89) & 94.77(0.41)  \\ \hline
        MTSDNet-A-tsg & 90.76(0.55) & 89.41(0.94) & 93.49(0.65) & 93.00(0.40)  \\ \hline
        MTSDNet-A-t3s3g3 & 91.03(0.95) & 89.59(1.11) & 93.01(0.92) & 93.20(0.76)  \\ \hline
        MTSDNet-M-tsg & 91.68(0.48) & 89.44(0.59) & 93.98(0.54) & 94.19(0.45)  \\ \hline
        MTSDNet-M-t3s3g3 & 91.77(0.35) & 89.12(0.36) & \textbf{94.59(0.38)} & 94.19(0.29)  \\ \hline
        DANN & 90.44(0.77) & 85.24(0.61) & 89.71(2.19) & 91.78(1.17)  \\ \hline
        GILE & 90.37(1.58) & \textbf{91.14(0.65)} & 93.03(1.00) & 94.53(0.63)  \\ \hline
    \end{tabular}
    }
\end{table}

Table~\ref{tab:table5} shows that all methods can gain the accuracy of over 90\% on the UCIHAR dataset, similar to the performance on the UniMib dataset, where the accuracy difference between the various methods is very small, with only about 3\% difference between the optimal and the worst average accuracy. Specifically, the GRU method achieved the best results on Part1 and Part4, GILE performed the best on Part2, and MTSDNet performed the best on Part3. 

\begin{table*}[ht]
\caption{The accuracy comparison of different methods on the PAMAP2 dataset from Part1 to Part8.\label{tab:table6}}
    \centering
    \resizebox{\textwidth}{!}{
    \begin{tabular}{|l|l|l|l|l|l|l|l|l|}
    \hline
        Method & Part1 & Part2 & Part3 & Part4 & Part5 & Part6 & Part7 & Part8  \\ \hline
        GRU & 64.63(1.59) & 55.11(2.11) & 74.96(2.02) & 69.71(3.04) & 72.18(1.20) & 70.05(1.98) & 76.21(3.08) & 36.81(3.58)  \\ \hline
        LSTM & 65.98(1.90) & 53.51(2.91) & 72.38(2.11) & 67.17(3.13) & 71.79(2.35) & 70.60(1.62) & 77.76(3.93) & 37.99(3.67)  \\ \hline
        MLP & 63.08(1.72) & 52.23(2.36) & 71.83(1.86) & 73.35(2.96) & 74.51(1.45) & 71.53(1.91) & 81.68(1.77) & 28.64(3.09)  \\ \hline
        Transformer & 66.84(1.14) & 59.37(2.32) & 73.60(2.53) & 72.64(3.83) & 75.52(1.73) & 74.58(1.92) & 75.11(4.93) & 31.93(5.62)  \\ \hline
        DeepConvLSTM & 63.79(2.63) & 64.28(2.19) & 75.88(2.23) & 72.37(2.48) & 75.05(1.36) & 74.98(2.87) & 77.60(2.31) & 54.96(5.49)  \\ \hline
        MTSDNet-A-tsg & 68.89(1.51) & 66.56(2.08) & 78.51(0.90) & 70.08(1.94) & 77.02(1.73) & 77.66(1.26) & 80.51(1.74) & 56.15(4.05)  \\ \hline
        MTSDNet-A-t3s3g3 & \textbf{70.53(1.05)} & 66.92(1.66) & \textbf{80.53(1.45)} & 71.11(1.33) & 77.15(1.30) & 78.92(1.42) & 80.85(2.03) & 60.72(1.76)  \\ \hline
        MTSDNet-M-tsg & 68.42(1.69) & 65.11(2.64) & 79.53(1.54) & 70.90(2.11) & 79.10(1.09) & \textbf{79.52(0.70)} & \textbf{82.97(1.44)} & 57.69(3.48)  \\ \hline
        MTSDNet-M-t3s3g3 & 68.54(1.71) & \textbf{66.93(2.47)} & 79.59(0.98) & \textbf{73.24(1.51)} & \textbf{80.27(1.27)} & 78.18(0.61) & 81.80(0.96) & 51.41(1.42)  \\ \hline
        DANN & 61.55(1.28) & 47.26(3.07) & 72.73(1.34) & 72.76(2.04) & 72.23(1.56) & 70.79(2.13) & 77.11(2.90) & 31.94(3.58)  \\ \hline
        GILE & 65.17(1.45) & 66.23(1.83) & 78.27(0.81) & 72.35(1.67) & 76.79(0.67) & 76.41(1.09) & 81.11(1.27) & \textbf{62.66(2.93)}  \\ \hline
    \end{tabular}
    }
\end{table*}

On the PAMAP2 dataset, the average classification accuracy of involved methods is about 67\% and there are significant volunteer differences of the average accuracy on Part2 and Part8 as shown in Table~\ref{tab:table6}. The classification accuracy of traditional methods under these two partitions is less than 60\%, while MTSDNet and GILE have significant accuracy improvements. Specifically, GILE performed the best on Part8 with the highest volunteer differences, and MTSDNet performed the best on Part2. MTSDNet achieved an average accuracy of 80\% on Part3, Part5, Part6 and Part7.

\begin{table}[ht]
\caption{The accuracy comparison of different methods on the OPPORTUNITY dataset from Part1 to Part4.\label{tab:table7}}
    \centering
    \resizebox{1\columnwidth}{!}{
    \begin{tabular}{|l|l|l|l|l|}
    \hline
        METHOD & Part1 & Part2 & Part3 & Part4  \\ \hline
        GRU & 48.29(5.81) & 57.70(8.64) & 32.63(5.82) & 37.68(6.94)  \\ \hline
        LSTM & 24.69(5.49) & 25.14(4.78) & 17.83(4.38) & 19.81(3.79)  \\ \hline
        MLP & 75.14(3.33) & 70.49(3.97) & 58.80(7.05) & 69.57(4.10)  \\ \hline
        Transformer & 50.94(6.88) & 56.2(13.05) & 35.52(11.01) & 47.24(12.98)  \\ \hline
        DeepConvLSTM & 56.45(26.11) & 61.48(5.21) & 43.83(4.34) & 57.17(20.55)  \\ \hline
        MTSDNet-A-tsg & \textbf{82.67(0.33)} & 78.97(1.05) & \textbf{71.89(4.88)} & 81.00(0.94)  \\ \hline
        MTSDNet-A-t3s3g3 & 81.24(6.18) & \textbf{79.50(1.09)} & 69.96(4.80) & \textbf{81.87(0.47)}  \\ \hline
        MTSDNet-M-tsg & 82.37(0.30) & 79.08(0.55) & 71.77(4.83) & 79.98(0.49)  \\ \hline
        MTSDNet-M-t3s3g3 & 82.37(0.29) & 78.74(0.67) & 56.64(24.31) & 80.59(0.41)  \\ \hline
        DANN & 57.25(8.38) & 64.91(4.74) & 47.47(10.82) & 52.34(8.63)  \\ \hline
        GILE & 60.09(10.20) & 67.66(7.76) & 49.41(6.32) & 56.11(7.61)  \\ \hline
    \end{tabular}
    }
\end{table}

As shown in Table~\ref{tab:table7}, MTSDNet exhibits significant advantages in both accuracy and stability on the OPPORTUNITY dataset, where the significant issue of label imbalance exists. On account of imbalanced labels, the accuracy of different methods varies greatly. LSTM and GRU perform poorly on this dataset, with the accuracy considerably lower than MLP, while GILE, Transformer and DeepConvLSTM have significant variances with classification accuracy ranging from over 80\% to around 20\%. In addition, MTSDNet-A performs better than MTSDNet-M. The data from MTSDNet-M-t3s3g3 shows that when using a multi-layer stacked multiplicative model on this dataset, there exists a notable model degradation phenomenon, with an average accuracy decrease of about 15\% and a significant improvement in accuracy variance.

Overall, MTSDNet has shown the improvements of prediction accuracy and stability in most cross-person HAR datasets, especially on the OPPORTUNITY and UniMib datasets. And the difference of prediction accuracy and stability between the additive model and the multiplicative model is small. The additive model can be applied to almost all situations.  Only in a few tasks like Part1 and Part3 on the DSADS dataset multiplicative model performs slightly better than additive model. The selection between additive and multiplicative models should be based on the data characteristics, which is similar to the traditional time series decomposition method.

\subsection{Ablation Experiment and Visualization}\label{Ablation}

\begin{table}[ht]
\caption{The ablation experiment on five datasets\label{tab:table8}}
    \centering
    \resizebox{1\columnwidth}{!}{
    \begin{tabular}{|l|l|l|l|l|l|}
    \hline
        Method & UCIHAR & PAMAP2 & DSADS & OPPORTUNITY & UniMib  \\ \hline
        MTSDNet-A-tsg & 91.67 & 71.92 & 91.28 & 78.63 & 98.92  \\ \hline
        MTSDNet-A-tsg-nostat & 86.97 & 69.60 & 80.33 & 74.27 & 96.90  \\ \hline
        MTSDNet-A-t3s3g3 & 91.70 & 73.34 & 92.16 & 78.14 & 98.85  \\ \hline
        MTSDNet-A-t3s3g3-nostat & 88.95 & 69.82 & 80.27 & 75.86 & 97.02  \\ \hline
        MTSDNet-M-tsg & 92.32 & 72.91 & 91.28 & 78.30 & 98.81  \\ \hline
        MTSDNet-M-tsg-nostat & 89.41 & 70.06 & 87.98 & 74.23 & 97.06  \\ \hline
        MTSDNet-M-t3s3g3 & 92.42 & 72.50 & 90.90 & 74.58 & 98.74  \\ \hline
        MTSDNet-M-t3s3g3-nostat & 91.37 & 69.52 & 89.10 & 77.35 & 97.32  \\ \hline
    \end{tabular}
    }
\end{table}

We further conducted the ablation experiments on whether to use statistical features. The suffix of '-nostat' shown in Table~\ref{tab:table8} indicates that the statistical features are not used in the models. The experiments focus on MTSDNet-A-tsg and MTSDNet-M-tsg, which are the most simplified form of MTSDNet. According to the results in Table~\ref{tab:table8}, the use of statistical features has a significant impact on the prediction accuracy, which indicates that the normalization of the sliding window can lose the features in the original signal. Therefore, during the construction of MTSDNet, the fusion of statistical features is indispensable. In addition, when statistical features are not used, the multiplicative model has a greater advantage than the additive model.

\begin{table}[ht]
\caption{The weight of MTSDNet-A-tsg and MTSDNet-A-tsg-nostat in the attention Layer.\label{tab:table9}}
    \centering
    \begin{tabular}{|l|l|l|l|l|}
    \hline
        DATASET & MODEL & T & S & G  \\ \hline
        UCIHAR & MTSDNet-A-tsg & 0.48 & 0.25 & 0.26  \\ \hline
        ~ & MTSDNet-A-tsg-nostat & 0.40 & 0.25 & 0.35  \\ \hline
        PAMAP2 & MTSDNet-A-tsg & 0.58 & 0.20 & 0.21  \\ \hline
        ~ & MTSDNet-A-tsg-nostat & 0.31 & 0.33 & 0.35  \\ \hline
        DSADS & MTSDNet-A-tsg & 0.62 & 0.21 & 0.17  \\ \hline
        ~ & MTSDNet-A-tsg-nostat & 0.44 & 0.27 & 0.28  \\ \hline
        UniMib & MTSDNet-A-tsg & 0.32 & 0.28 & 0.39  \\ \hline
        ~ & MTSDNet-A-tsg-nostat & 0.37 & 0.27 & 0.34  \\ \hline
    \end{tabular}
\end{table}

 In order to further investigate the preference of MTSDNet for statistical features, the feature values of layer attention are analyzed and the results are presented in Table~\ref{tab:table9}. MTSDNet-A-tsg is used for the convenience of experimental comparison. As seem from Table~\ref{tab:table9}, when using statistical features, the layer attention of the first layer has significantly improved on most datasets. On PAMAP2, DSADS and OPPORTUNITY datasets with significant differences in individual activity, the attention weights are increased by 0.1 to 0.2, respectively. This indicates that compared to structures without statistical features, the model attaches more importance to the statistical information of original signal for final classification. On the UCIHAR and UniMib datasets with little difference in individual activity, the increase of weight value is not significant in the first layer attention. The mentioned phenomenon suggests that in the presence of substantial volunteer variability, statistical features of the original signal exhibit stronger generalization capabilities compared to other feature types. This aligns with traditional practices of feature engineering during the feature design. Conversely, when inter-volunteer differences are relatively minor, the statistical properties of the original signal may not receive sufficient consideration. 

\begin{figure*}[!t]
\centering
\includegraphics[width=5.5in]{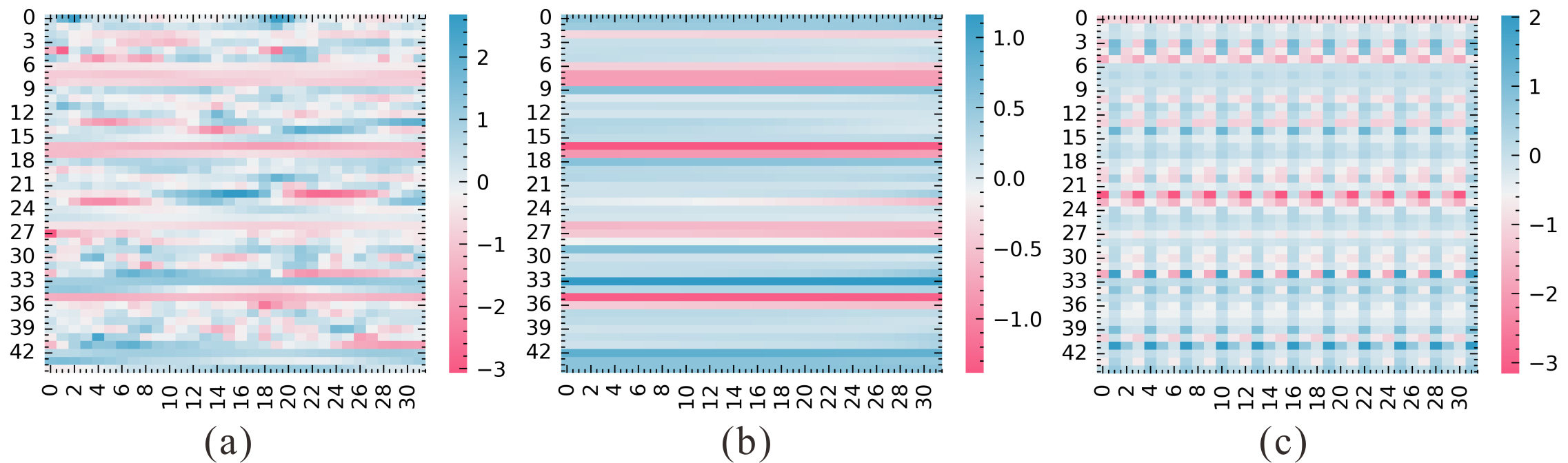}
\caption{The time series of a sample from DSADS dataset is decomposed into multiple components for the visualization of MTSDNet-A-tsg model. (a) Original signal. (b) Trend component. (c) Seasonal component.}
\label{Visualization_component}
\end{figure*}

\begin{figure*}[!t]
\centering
\includegraphics[width=5.5in]{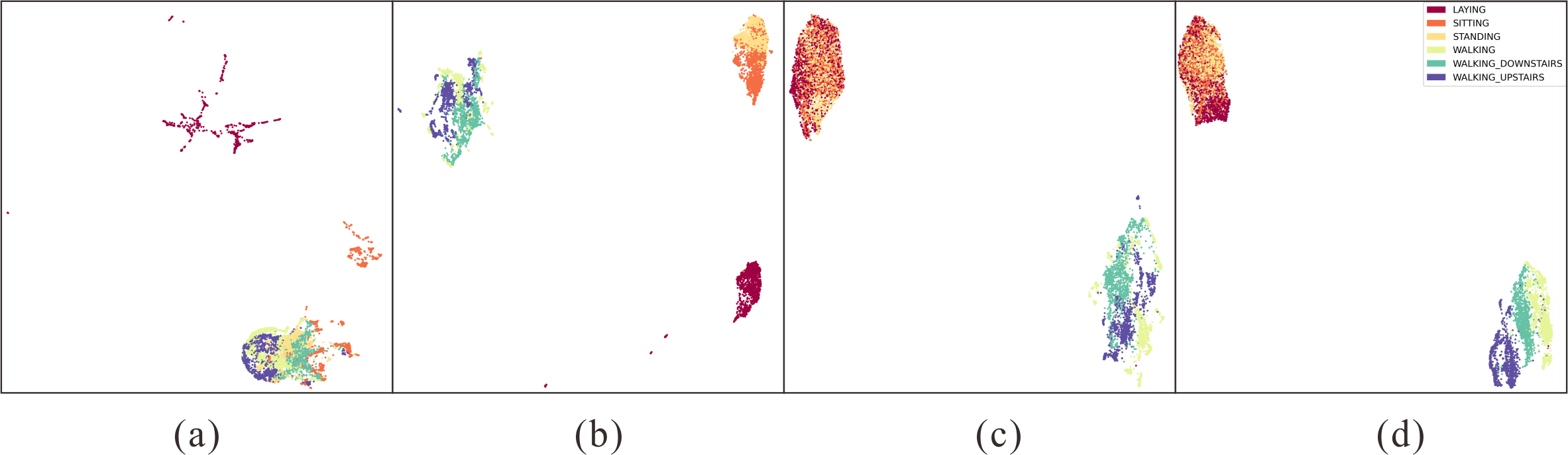}
\caption{Visualization of high-dimensional features from UCIHAR dataset using UMAP for MTSDNet-A-tsg model. (a) Original signal. (b) Trend component. (c) Seasonal component. (d) General component.}
\label{Visualization_UCIHAR}
\end{figure*}

\begin{figure*}[!t]
\centering
\includegraphics[width=5.5in]{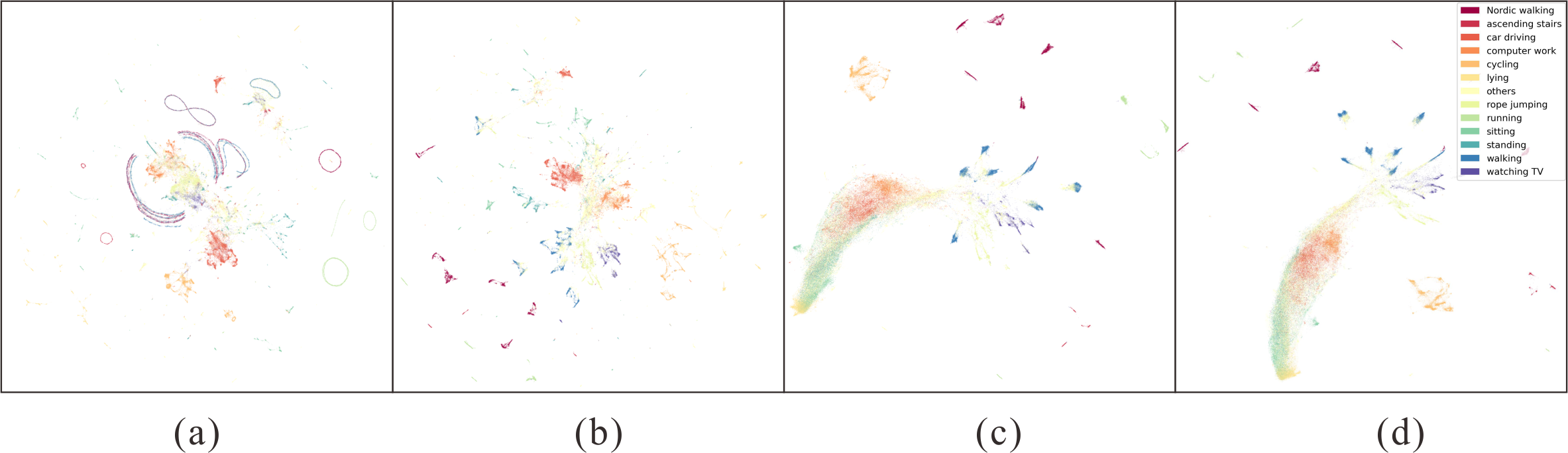}
\caption{Visualization of high-dimensional features from PAMAP2 dataset using UMAP for MTSDNet-A-tsg model. (a) Original signal. (b) Trend component. (c) Seasonal component. (d) General component.}
\label{Visualization_PAMAP2}
\end{figure*}

In addition, we visualize the extracted signals from temporal decomposition and the characteristics of decomposition signals at different layers to show the interpretability and distribution differences of MTSDNet-A-tsg model at each layer. Figure~\ref{Visualization_component} shows the temporal decomposition visualization of a single sample on the DSADS dataset. 
From Figure~\ref{Visualization_component} (b) and (c), the signal extracted by decomposer has a low-rank characteristic. Trend term in Figure~\ref{Visualization_component} (b) can be approximated as the mean value extracted to the input signal. The polynomial constraint has high coefficients only on the constant term, with the rest of the coefficients close to zero. Seasonal term in Figure~\ref{Visualization_component} (c) has a clear preference for a certain frequency and it can be seen that seasonal  term is biased towards high-frequency signals. We believe that there may be two reasons: firstly, the size of sliding window currently used is difficult to extract low-frequency signals; Secondly, similar to traditional time series decomposition methods, when the periodicity of the original signal is not remarkable, the seasonal term is easily extracted from high-frequency signals. Owing to the insignificant effect of high-frequency signals on classification, the model has assigned lower weights to seasonal layer, as shown in Table~\ref{tab:table9}.

Furthermore, by adopting the UMAP~\cite{RN13} method, we visualize the high-dimensional features in different decomposition layers of MTSDNet-A-tsg model for input data. Figure~\ref{Visualization_UCIHAR} and Figure~\ref{Visualization_PAMAP2} are the visualization results on the UCIHAR and PAMAP2 datasets. The reason for using UCIHAR and PAMAP2 datasets is that UCIHAR dataset has weak differences in distribution while PAMAP2 dataset is just the opposite, thus making the visualization diverse. The visualization is repainted with different colors for every activity. UMAP can effectively reflect the correlation between data and represent it by the distance between data points. UMAP can even exhibit periodic characteristics of data by visualizing it as linear or circular shapes, which can be seen in Figure~\ref{Visualization_PAMAP2} (a). The visualization of high-dimensional features from UCIHAR dataset using UMAP for MTSDNet-A-tsg model is shown in Figure~\ref{Visualization_UCIHAR}, which indicates that the trend term effectively separates static activities (laying, sitting, standing), while the seasonal term and general term are almost identical from static activities and focus more on the division of dynamic activities (walking, upstairs, downstairs). 
This means that different levels of attention gain diversity in the focused features. The distribution difference on the UCIHAR dataset is small, so from the visualization we can see that the same activity from different volunteers are clustered in one group.

PAMAP2 dataset has a significant difference of data distribution, as shown in Figure~\ref{Visualization_PAMAP2}, which is relatively scattered and chaotic. From Figure~\ref{Visualization_PAMAP2}~(a) to Figure~\ref{Visualization_PAMAP2}~(c), the scattered points are clustered into fewer and more dense clusters. For activities such as going up and down stairs, standing and sitting, it shows that there are obvious distribution states of multiple clusters, which correspond to different volunteers performing the same activity. 
As shown in Figure~\ref{Visualization_PAMAP2}~(c) and Figure~\ref{Visualization_PAMAP2}~(d), the differences in some activities are eliminated, and multiple volunteers either gather together in the form of point clouds (green, yellow and orange) or are in adjacent areas (crimson, blue and purple) with the form of multiple clusters. 
From the global positions of the clusters, the features obtained from temporal decomposition alleviate differences among volunteers and highlight different activities.

The visual structure of seasonal and general terms as shown in Figure~\ref{Visualization_UCIHAR}~(c,d) and Figure~\ref{Visualization_PAMAP2}~(c,d) is very similar, implying that the extracted features are homogeneous. But there are slight differences in the global structure and the distance arrangement between some clusters has changed.
The above results are consistent with the general cognition, which reflects the reliability of the constructed model. Furthermore, it can be observed that the clusters of general terms exhibit a higher degree of spatial clustering in comparison to seasonal terms. 

In addition, there was no significant difference between MTSDNet-A and MTSDNet-M in all the experiments mentioned above. This lack of differentiation can be attributed to the absence of multiplicative interference in HAR, rendering MTSDNet-M, based on multiplicative models, without evident advantages. In time series signals with additive interference, both can express signal characteristics well, and MTSDNet-A is suitable for most tasks.

Overall, by decomposing the original signal into multiple components, we can find remarkable clustering in the visualization results, indicating the difference reduction between volunteers and simultaneously emphasizing their behavioral patterns. The clustering degree of visualization suggests that trend term displays greater structural variability in contrast with both seasonal term and general term, thus demonstrating more notable discriminatory power of its feature extraction. In other words, the characteristic structures of the seasonal and general terms are more homogeneous.

\section{Conclusion}
This study analyzes the signal characteristics of sensor-based cross-person HAR tasks and investigates domain-specific features called data bias and volunteer activity differences. To process components separately, this work presents MTSDNet, a novel sensor data-based method for cross-person HAR task with the ability of temporal decomposition. MTSDNet decomposes the original signal into low-rank representations of polynomials and trigonometric functions, enabling the model to learn more generalized features and enhancing its domain generalization ability. MTSDNet does not require additional information and is suitable for the prediction of new domain. The experiments on five different HAR public datasets demonstrate that MTSDNet gains the average accuracy improvement of 9.2\%, 28\%, 4.2\%, 6.8\% and 1.3\% on DSADS, OPPORTUNITY, UniMib, PAMAP2 and UCIHAR datasets, respectively and performs good stability in most cases. 

In addition, the visualization demonstrates that MTSDNet extracts differentiated features through temporal decomposition, which indirectly reflects the effectiveness and interpretability of temporal decomposition in domain generalization. By the idea of decomposing original signal into multiple components to reduce distribution differences, the performance of domain generalization is effectively improved, which can also extend to other applications including remote sensing images, where spatio-temporal disparities caused by the factors like fog, season and lighting introduce the data bias. 

\bibliographystyle{IEEEtran}
\bibliography{IEEEabrv,references}

\newpage
\vspace{11pt}



\vfill

\end{document}